\documentclass[aps,graphicx,twocolumn,epsfig]{revtex4}

\usepackage{epsfig}

\begin{document}

\title{Pair-supersolid phase in a bilayer system of dipolar lattice bosons}

\author{C. Trefzger $^1$}
\author{C. Menotti $^2$}
\author{M. Lewenstein $^{1,3}$}
\affiliation{$^1$ ICFO - Institut de Ciencies Fotoniques,
Mediterranean Technology Park, 08860 Castelldefels (Barcelona), Spain \\
$^2$ CNR-INFM BEC and Dipartimento di Fisica, Universit\`a di
Trento, I-38050 Povo, Italy\\
$^3$ ICREA and  ICFO - Institut de Ciencies Fotoniques,
Mediterranean Technology Park, 08860 Castelldefels (Barcelona),
Spain}

\begin{abstract}
The competition between tunneling and interactions in bosonic
lattice models generates a whole variety of different quantum
phases. While, in the presence of a single species interacting via
on-site interaction, the phase diagram presents only superfluid or
Mott insulating phases, for long-range interactions or multiple
species, exotic phases such as supersolid (SS) or pair-superfluid
(PSF) appear. In this work, we show for the first time that the
co-existence of effective multiple species and long-range
interactions leads to the formation of a novel pair-supersolid
(PSS) phase, namely a supersolid of composites. We propose a
possible implementation with dipolar bosons in a bilayer
two-dimensional optical lattice.
\end{abstract}

\maketitle

The possibility of engineering lattice models with ultra-cold
gases in optical lattices is considered one of the most promising
routes in the search for exotic quantum phases which escape clean
demonstration in condensed matter systems. To the aim of the
present work, where we show the existence of a pair-supersolid
(PSS) phase, it is particularly important to briefly introduce the
supersolid (SS) and the pair-superfluid (PSF) phases.

The question whether superfluidity and broken translational
symmetry can coexist, leading to supersolidity, has been
intriguing theoretical and experimental physicists for the last 50
years (see e.g. \cite{prokofev2008,kimchan,pollet2007,svistunov2009,sasaki2006,ray2008}). 
%The existence of the
%supersolid phase in solid He-4 has not yet been unambiguously
%proven experimentally: while on the one hand the interpretation of
%the first experimental results measuring a non-classical
%rotational inertia \cite{kimchan} remains controversial,
%microscopic calculations \cite{pollet2007,svistunov2009} indicate
%that disorder-based mechanisms should be responsible for the more
%recent observations of supersolidity \cite{sasaki2006,ray2008}.
Exact Quantum Monte Carlo simulations have demonstrated the
possibility of a SS ground state in lattices
 \cite{batrouni2000,hebert2001,sengupta2005,chen2007}.
%Exact Quantum Monte Carlo simulations have lead to an agreement
%concerning the possibility of having a checkerboard SS phase in
%lattices for incommensurate densities and strong enough nearest
%neighbor interaction
%\cite{batrouni2000,hebert2001,sengupta2005,chen2007}.
 The
experimental realization of a strong dipolar Chromium condensate
\cite{lahaye2007} and the recent progresses towards a degenerate
gas of heteronuclear polar molecules
\cite{danzl2008,ni2008,ospelkaus2008} put cold gases with
long-range interaction in optical lattices among the best
candidates for the creation of the SS phase \cite{goral2002,
kovrizhin2005, scarola2006}.

The second important issue concerns particle vs pair condensation
in Bose gases in presence of attractive interactions
\cite{nozieres1982}. It has been shown %by mean-field calculations
%\cite{altman2003}, confirmed by Quantum Monte Carlo
%\cite{kuklov2003, kuklov2004, soyler2008} and dynamical mean-field
%simulations \cite{hubener2009}
\cite{altman2003,kuklov2003, kuklov2004, soyler2008,hubener2009,paredes2003}, that bosonic mixtures with
inter-species attraction can actually sustain a pair-superfluid
phase (PSF) without either collapsing or phase separating. %The
%analogy with fermionic pairing has been discussed in
%\cite{paredes2003}.
The optimal candidates for the realization of
the PSF phase are bosonic binary mixtures in optical lattices with
interspecies interactions tunable via a Feshbach resonance
\cite{catani2008} or, alternatively, bilayer optical lattices of
dipolar particles, which create an effectively two-species system
if tunneling between the two layers is suppressed \cite{arguelles2007}.
%, since the particles
%belonging to the different layers cannot mix and behave in
%practice like two different species.
The "inter-species" interaction is provided by the long-range
interaction, which couples the two separate sub-systems. 
%The
%existence of the PSF phase has been recently demonstrated in a
%system of dipoles in two one-dimensional (1D) optical lattices,
%for an appropriate choice of the polarization of the dipoles such
%to produce no intra-wire nearest neighbor interaction
%\cite{arguelles2007}.

However, dipolar gases offer the further opportunity of studying
the combined effect of long-range interactions and inter-species
coupling. In this paper, we show for the first time that the
presence of both intra-layer repulsion and inter-layer attraction
allows for a pair-supersolid phase (PSS), defined as a supersolid
phase of composites. This phase, which joins the characteristics
of the supersolid phase and the pair-superfluid phase, can be
obtained in a system of dipolar bosons populating two decoupled
two-dimensional (2D) optical lattice layers.

\begin{center}
\begin{figure}[b]
\includegraphics[width=0.7\linewidth]{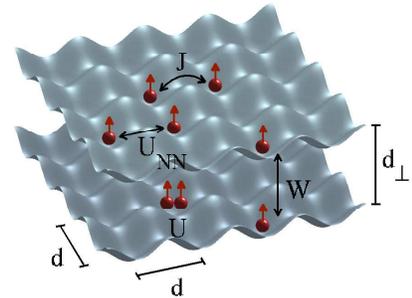}
\caption{Schematic representation of two 2D optical lattice layers
populated with dipolar bosons polarized perpendicularly to the
lattice plane. The particles feel  repulsive on-site $U$ and
nearest neighbor $U_{\rm NN}$ interactions. Inter-layer tunneling
is completely suppressed, while a  nearest neighbor inter-layer
attractive  interaction $W$ is present.} \label{fig_system}
\end{figure}
\end{center}

%%% REMOVE SPACING AFTER FIGURES!!
%\vspace*{-0.8cm}

 In order to demonstrate the existence of the PSS,
we study the effective Hamiltonian $H_{\rm eff}$ in the low-energy
subspace of pairs, using a mean-field Gutzwiller approach and
exact diagonalization. Moreover, there exist an accurate mapping
at low densities of the effective Hamiltonian onto the single
component extended Hubbard model \cite{sengupta2005}; the fact
that the latter model supports  SS implies the existence of PSS in
our case.

We consider polarized dipolar particles in two decoupled 2D layers
(see Fig.~\ref{fig_system}). This geometry can be obtained by
using anisotropic optical lattices or superlattices, which can
exponentially suppress tunneling in one direction. The in-plane
dipolar interaction is isotropic and repulsive. The inter-layer
interaction depends on the relative position between the two
dipoles, but is dominated by the nearest neighbor attractive
interaction $W<0$ between two atoms at the same lattice site in
different layers. In the following, we include only nearest
neighbor in-plane ($U_{\rm NN}$) and out-of-plane ($W$) dipolar
interactions. The relative strength between $U_{\rm NN}$ and $W$
can be tuned by changing the spacing $d_{\perp}$ between the two
layers, relative to the 2D optical lattice spacing $d$. Due to the
dependence of the dipole-dipole interaction like the inverse cubic
power of the distance, the ratio $|W|/U_{\rm NN}$ can be tuned
over a wide range. While it can be negligible for $d_{\perp} \gg
d$ making the system asymptotically similar to a single 2D lattice
layer, it can also become relevant and give rise to interesting
physics, not existing in the single layer model
\cite{arguelles2007, chen2003, wang2006, wang2007, yi2007,
wang2008, klawunn2008}.
%In particular, in the case considered
%here, the interplay between intra-species long-range repulsion and
%inter-species attraction leads to the formation of the PSS phase,
%a supersolid phase of pairs.

We start from the generalized extended Bose-Hubbard Hamiltonian

\begin{eqnarray}
H&=&\sum_{i,\sigma} \left[\frac{U}{2} n_i^\sigma (n_i^\sigma -1) +
\sum_{\langle j\rangle_i} \frac{U_{\rm NN}}{2} n_i^\sigma
n_j^\sigma -  \mu n_i^\sigma \right] + \nonumber \\
&& + \sum_i W n_i^a n_i^b- J \sum_{\langle ij \rangle}
\left[a_i^\dag a_j + b_i^\dag b_j \right], \label{h}
\end{eqnarray}
where $\sigma=a,b$ indicates the two species (which in the
specific case considered here are atoms in the lower and upper 2D
optical lattice layer respectively), $U$ is the on-site energy,
$U_{\rm NN}$ the intra-layer nearest neighbors repulsion, $W$ the
inter-layer attraction, $J$ the intra-layer tunneling parameter
and $\mu$ the chemical potential. The parameters $U$ and $J$ are
equal for the upper and lower layers and the chemical potentials
$\mu$ are the same, since equal densities in the two layers are
assumed. The symbols $\langle j\rangle_i$ and $\langle ij \rangle$
indicate nearest neighbors.

The PSS is characterized by vanishing single particle order
parameters $\langle a \rangle = \langle b \rangle =0$, and non
vanishing pair order parameter $\langle a b \rangle \neq 0$,
coexisting with broken translational symmetry, namely a modulation
of both density and order parameter on a scale larger than the one
of the lattice spacing, analogously to the supersolid phase. The
physics leading to the formation of composites relies on second
order tunneling and takes place in the low-energy subspace where
single-particle hopping is suppressed. The
theoretical description of the PSS phase cannot be based on standard
mean-field theory, which accounts for particle hopping through the
replacement in the Hamiltonian of the single particle creation and
destruction operators by their expectation values, because in this approximation
second order tunneling completly vanishes.

A successful way to account for second order tunneling is to write
an effective Hamiltonian in the subspace of pairs and include
tunneling through second order perturbation theory
\cite{altman2003,kuklov2003,kuklov2004,cct}. The validity of the
effective Hamiltonian relies on the existence of a low-energy
subspace well separated in energy from the subspace of virtual
excitations, to which it is coupled via single particle hopping.
Such second order couplings are related to the super-exchange
interaction, recently measured in a series of double-well systems
\cite{trotzky2008}.

\begin{center}
\begin{figure}[t]
\includegraphics[width=0.85\linewidth]{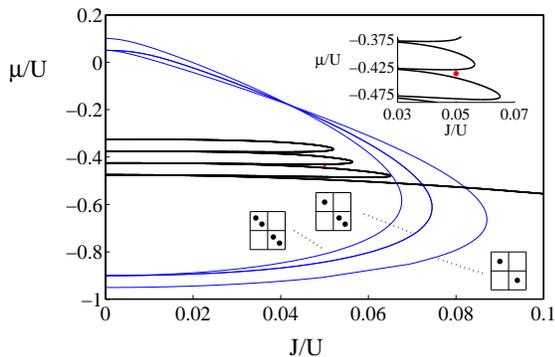}
\caption{Pair insulating lobes for $\nu=0,1/2,1,3/2$ (thick
lines); Lobes with respect to single particle-hole excitations
(thin blue lines) for the dominant configurations in the ground
state at $J=0.05U$ and $\mu=-0.4375U$, namely $m_i=0$ and $m_j=1,2$
(for $i,j$ nearest neighboring sites). The inset shows a zoom of
the pair phase diagram.} \label{validity}
\end{figure}
\end{center}

The low-energy subspace of pairs is spanned by all classical
distributions of atoms in the lattice $|\alpha\rangle=\prod_i
|n_i,n_i\rangle$ with equal occupation of the two species $a$ and
$b$.
For $(U+W)/U \to 0$, asymptotically all classical states
$|\alpha\rangle$ become stable with respect to single
particle-hole excitations and develop an insulating lobe at finite
$J$. The energy of single particle-hole excitations is of the
order of $U$ at $J=0$ and is given by the width of the lobes at
finite $J$ (see e.g. thin blue lobes in Fig.\ref{validity}). This
situation has to be compared to the single layer situation ($W=0$)
with nearest neighbor interactions, where only uniform Mott phases
and checkerboard insulating phases are stable.
The relevant virtual subspace is obtained from
the states $|\alpha\rangle$ by breaking one composite, namely
$|\gamma_{ij}^{(a)} \rangle = a_i^\dag a_j |\alpha\rangle
/\sqrt{n_j(n_i+1)}$ and $|\gamma_{ij}^{(b)} \rangle =b_i^\dag b_j
|\alpha\rangle /\sqrt{n_j(n_i+1)}$.
%All other states are not
%coupled to $|\alpha\rangle$ via single particle hopping and hence
%do not contribute to second order tunneling.

In the pair-state basis, the matrix elements of the Hamiltonian in
second order perturbation theory are given by

\begin{eqnarray}
&&\langle \alpha |H_{\rm eff}|\beta \rangle= \langle \alpha
|H_{0}|\beta \rangle +\\
&&- \frac{1}{2} \sum_{\gamma} \langle \alpha |H_{1}|\gamma \rangle
\langle \gamma |H_{1}|\beta \rangle  \left[
\frac{1}{E_{\gamma}-E_{\alpha}} + \frac{1}{E_{\gamma}-E_{\beta}}
\right], \nonumber
\end{eqnarray}
where $H_0$, given by the interaction terms of the Hamiltonian
(\ref{h}), is diagonal on the states $|\alpha\rangle$, and the
single particle tunneling term $H_1=- J \sum_{\langle ij \rangle}
\left[a_i^\dag a_j + b_i^\dag b_j \right]$ is treated at second
order.

For a given state $|\alpha\rangle$,

\begin{eqnarray}
E_{\gamma_{ij}}-E_\alpha=U+(U+W)(m_i-m_j)+ U_{\rm NN} \Delta
m^{ij}_{\rm NN}, \label{den}
\end{eqnarray}
with $\Delta m^{ij}_{\rm NN}=\sum_{\langle
k\rangle_i}m_{k}-\sum_{\langle k \rangle_j} m_{k}-1$, where $m_i$
indicates the pair occupation number at site $i$. For
$\mbox{U+W},U_{\rm NN}\ll U$, the denominators
$E_{\gamma_{ij}}-E_\alpha$ are all of order $U$, which leads to

\begin{eqnarray}
\label{heff0} &&H_{\rm eff}^{(0)}=H_0 - \frac{2J^2}{U} \sum_{\langle ij
\rangle} \left[ m_i (m_j+1) + c_i^\dag c_j \right],
\end{eqnarray}
where $c_i$ and $c_i^\dag$ are the pair destruction and creation
operators such that $c_i^\dag|m_i\rangle=(m_i+1)|m_i\rangle$.  One
can easily obtain corrections to $H_{\rm eff}^{(0)}$  by expanding
(\ref{den}) at higher orders in $(U+W)/U$ and $U_{\rm NN}/U$.

In this work, we provide a mean-field solution to effective
Hamiltonian (\ref{heff0}). We perform a perturbative treatment at
first order in the pair order parameter $\psi = \langle c \rangle$
to determine the boundaries of the insulating lobes. Furthermore,
we solve the time dependent Gutzwiller equations in imaginary time
to determine the nature of the superfluid phases outside the
insulating lobes. We point out that in spite of its simplicity,
the mean-field treatment of the effective Hamiltonian is able to
include important features like the re-entrant behavior of the
lobes (as predicted by exact t-DMRG calculations for the 1D
geometry in \cite{arguelles2007}).

\begin{center}
\begin{figure}[t]
\includegraphics[width=0.85\linewidth]{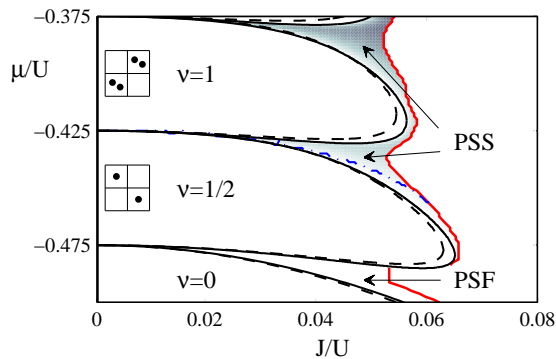}
\caption{Phase diagram of the effective Hamiltonian, with $U_{\rm
NN}=0.025U$, $W = -0.95 U$, which can be obtained for
$d_{\perp}=0.37d$. The black full lines are the semi-analytic
solution of Eq.(\ref{Eqn:MF}) indicating the boundaries of the
insulating lobes for the checkerboard ($\nu=1/2$) and the doubly
occupied checkerboard ($\nu=1$). The black dashed lines are the
boundaries of the insulating lobes for 1st order expansion of
$H_{\rm eff}$. The shaded area is the PSS phase predicted by the
Gutzwiller approach. The red line indicates the estimated limit of
validity of $H_{\rm eff}^{(0)}$. The blue dash-dotted line
indicates the upper limit of the region where the mapping onto
\cite{sengupta2005} is almost exact.} \label{phase_diag}
\end{figure}
\end{center}

For any non-zero $W<0$, the lowest possible excitations on top of
a classical configuration of pairs are obtained by adding
(removing) one pair at site $i$. The energy costs, respectively
given by $E_{\rm P}^i(J) = -2\mu + 2U m_i + (2m_i + 1)W + 2 V_{\rm
dip}^{1,i}
           - (2J^2/U)\sum_{\langle k \rangle_i}(2m_k + 1)$
and $E_{\rm H}^i(J) =  2\mu - 2U(m_i-1) - (2m_i - 1)W - 2 V_{\rm
dip}^{1,i}
           + (2J^2/U)\sum_{\langle k \rangle_i}(2m_k + 1)$,
depend quadratically on the tunneling coefficient $J$. In the
previous expressions, $V_{\rm dip}^{1,i} > 0$ is the dipole-dipole
interaction of one atom placed at site $i$ with the rest of the
particles belonging to the same layer. The mean-field  order
parameters $\psi_i$ satisfy

\begin{equation}
\label{Eqn:MF}
 \psi_i = \frac{2J^2}{U}\left[ \frac{(m_i+1)^2}{E_{\rm P}^i(J)}
 + \frac{m_i^2}{E_{\rm H}^i(J)} \right]\bar{\psi_i},
\end{equation}
with $\bar{\psi_i} = \sum_{\langle k\rangle_i} \psi_k$ being the
sum of the nearest-neighbour order parameters. Using
Eq.~(\ref{Eqn:MF}), one can calculate the mean-field lobes for any
given configuration of pairs in the lattice.
The lobes for the checkerboard and doubly occupied checkerboard
are shown in Fig.~\ref{phase_diag} for the 0th (full lines) and
1st order (dashed lines) effective Hamiltonians.  The comparison
between the two shows that, for the parameters considered here,
the 0th order already captures the physics accurately. The $J^2$
dependence of the energy of the elementary excitations is at the
origin of the re-entrant behavior of the lobes.

%%% REMOVE SPACING AFTER FIGURES!!
%\vspace*{-0.8cm}

Based on the Gutzwiller Ansatz for the pair wavefunction
$|\Phi\rangle= \prod_i \sum_m f_m^{(i)} |m,i\rangle$, we predict
the existence of three different phases: insulating phases, PSF
and PSS. For $U+W<zU_{\rm NN}$, being $z$ the number of nearest
neighbors, the insulating phases show checkerboard ordering not
only at filling factor $\nu=1/2$, but also at filling factor
$\nu=1$. Outside the insulating lobes, depending on density and
tunneling, we find either PSF or PSS. The stability analysis of
the PSS phase against phase separation is beyond our mean-field
treatment. However,  for small $U + W<zU_{\rm NN}$, by doping the
checkerboard above half filling, the extra pair goes on an already
occupied site and the analogy to the single-species extended
Bose-Hubbard model \cite{sengupta2005} suggests that the system
stabilizes to a PSS phase. Instead, the PSS phase at density lower
than $1/2$ (which our mean-field treatment predicts only in a very
small region close to the tip of the $\nu=1/2$ lobe) should be
unstable towards phase separation. Nevertheless, the inclusion of
more neighbors to the in-plane dipolar interaction is expected to
remove this instability \cite{capogrosso2009}.

To get reliable results, one should combine the Gutzwiller
predictions with an estimate of the limits of validity of $H_{\rm
eff}^{(0)}$, beyond which the subspace of pairs looses its
meaning. For each point of the phase diagram, from the ground
state Gutzwiller wavefunction, we select the dominant classical
configurations with the criteria $|f_m^{(i)}|^2>0.05$ and
$|\prod_i f_m^{(i)}|>0.02$ \cite{footnote} and calculate for each
of these configurations, the lobe with respect to single
particle-hole excitations. If the system at this given point of
the phase diagram turns out to be stable against all dominant
single particle-hole excitations (in other words, if this point is
inside all selected single particle-hole lobes), $H_{\rm
eff}^{(0)}$ is considered valid at that point. This procedure is
shown for $J=0.05U$ and $\mu=-0.4375U$ in Fig.~\ref{validity}. In
Fig.~\ref{phase_diag}, we show the resulting phase diagram for
$U_{\rm NN}=0.025U$ and $W=-0.95 U$. The shaded area represents
the region of PSS, compatible with the above validity conditions
for $H_{\rm eff}^{(0)}$. Interestingly, the \mbox{$\prec$-like}
shape of the allowed pairing regions (PSF and PSS) matches the
ones found in \cite{arguelles2007}.

%%% REMOVE SPACING AFTER FIGURES!!
%\vspace*{-0.8cm}

The existence of the PSS phase is supported also by the exact
diagonalization of  $H_{\rm eff}^{(0)}$ for up to 6 pairs on a
$2\times 8$ lattice with periodic boundary conditions. At small
$J$, the ground state is almost doubly degenerate. This can
correspond to two insulating checkerboards (for half filling) or
two checkerboard supersolids shifted by one lattice constant, with
a large gap to excited states. This double degeneracy, together
with a finite measure of the coherence, provided by a non
vanishing expectation value of the tunneling term, are indications
of the existence of the PSS. At larger $J$, the quasi-degeneracy
of the two lowest eigenstates disappears and the ground state
becomes well separated in energy from all excited states,
indicating the crossover to the PSF. It is important to remark
that in the absence of nearest-neighbor interaction, the ground
state is always non degenerate and no signature for the PSS is
ever found.

Finally, we observe that, upon appropriate renaming of the
parameters, $H_{\rm eff}^{(0)}$  (\ref{heff0}) can be mapped onto
the Hamiltonian used in \cite{sengupta2005} to demonstrate the
existence of the SS phase for soft core bosons with nearest
neighbor interactions. Due to the different action of the
tunneling terms, namely $\langle
n_i+1,n_j-1|a^{\dag}_ia_j|n_i,n_j\rangle = \sqrt{(n_i+1)n_j}$ in
\cite{sengupta2005} and $\langle m_i+1,m_j-1|c^{\dag}_i
c_j|m_i,m_j\rangle = (m_i+1)m_j$ in our case, the mapping of the
two Hamiltonians is exact when only the number states $m=0,1$ are
populated. Under this condition, the results of
\cite{sengupta2005} translate to the existence of the PSS in our
problem. The region below the dash-dotted blue line in
Fig.~\ref{phase_diag}, which corresponds to \mbox{$|f_0^{(i)}|^2 +
|f_1^{(i)}|^2 > 0.9$}, $\forall i$, namely a close to exact
mapping, includes part of the PSS region.

Summarizing, we have studied the phase diagram of a bilayer system
of 2D dipolar lattice gases, in the limit of close layers, and
demonstrated the existence of a novel PSS phase, namely a
supersolid phase of pairs. The existence of the PSS phase has been
previously discussed for anisotropic $t$-$J$ models
\cite{boninsegni2008}, but no evidence of it has been found.
However, the Hamiltonian we discuss in the present work differs
from the anisotropic $t$-$J$ spin Hamiltonian in three crucial
respects, all of which should favor the existence of the PSS
phase: (i) it deals with soft-core bosons (vs. hard-core); (ii) it
considers on-site inter-species attraction (vs. nearest-neighbor
inter-species attraction); (iii) it includes nearest-neighbor
intra-species repulsion. For these reasons, we believe that the
existence of the PSS phase will be confirmed by exact
calculations, also beyond the limits of validity of our effective
mean-field approach.

We acknowledge support of Spanish MEC (FIS2008-00784, QOIT), EU
projects  SCALA and NAMEQUAM, ERC grant QUAGATUA.  The authors
thank  O. Dutta, M. Modugno, L. Santos, S. Stringari, B.
Svistunov, for interesting discussions.


\begin{thebibliography}{99}
\bibitem{footnote} We have checked that the validity region is not
strongly modified upon small changes in these conditions.
\end{thebibliography}
\end{document}